\documentclass[letterpaper, 10 pt, conference]{ieeeconf}  

\IEEEoverridecommandlockouts                              

\usepackage{graphicx}
\usepackage{subcaption}
\usepackage{algorithm}
\usepackage{algpseudocode}
\usepackage{amsmath}
\usepackage{tabularx}

\usepackage{tikz}
\usepackage{textcomp}
\usepackage{hyperref}
\usepackage{lipsum}

\newcommand\copyrighttext{%
  \footnotesize \textcopyright 2024 IEEE. Personal use of this material is permitted.
  Permission from IEEE must be obtained for all other uses, in any current or future
  media, including reprinting/republishing this material for advertising or promotional
  purposes, creating new collective works, for resale or redistribution to servers or
  lists, or reuse of any copyrighted component of this work in other works.}
\newcommand\copyrightnotice{%
\begin{tikzpicture}[remember picture,overlay]
\node[anchor=south,yshift=10pt] at (current page.south) {\fbox{\parbox{\dimexpr\textwidth-\fboxsep-\fboxrule\relax}{\copyrighttext}}};
\end{tikzpicture}%
}

\title{\LARGE \bf
An Application Layer Multi-Hop Collective Perception Service for Vehicular Adhoc Networks
}

\author{Vincent Albert Wolff$^{1}$, Edmir Xhoxhi$^{1}$ and Felix Tautz$^{1}$ 
\thanks{$^{1}$The authors are with Institute of Communications Technology,
        Leibniz University Hannover, Germany
        {\tt\small forename.lastname@ikt.uni-hannover.de}}%
}

\begin{document}

\maketitle
\copyrightnotice
\thispagestyle{empty}
\pagestyle{empty}

\begin{abstract}

Collective Perception will play a crucial role for ensuring vehicular safety in the near future, enabling the sharing of local perceived objects with other Intelligent Transport System Stations (ITS-Ss). However, at the beginning of the roll-out, low market penetration rates are expected. This paper proposes and evaluates an application layer multi-hop Collective Perception Service (CPS) for vehicular ad-hoc networks. The goal is to improve the environmental awareness ratio in scenarios with low CPS market penetration. In such scenarios, the CPS service without forwarding enabled struggles to achieve complete awareness. A decentralized application layer forwarding algorithm is presented that shares perceived object information across multiple hops while maintaining a low age of information. The proposed approach is compared against standard CPS with no forwarding and CPS with geographically-scoped (GBC) multi-hop forwarding. Simulations according to standards of the European Telecommunications Standards Institute (ETSI) demonstrate that the application layer forwarding achieves near 100\% awareness at 10\% penetration rate versus 92\% for standard CPS. The awareness improvement comes with moderate channel load, unlike GBC forwarding which quickly saturates the channel. The median age of information remains below 80 ms for the proposed scheme, enabling real-time CPS operation. Our application layer multi-hop approach effectively improves environmental awareness during initial CPS deployment while aligning with latency and channel load requirements.

\end{abstract}

\section{INTRODUCTION}
In recent years, the landscape of vehicular technology has witnessed a significant transformation, particularly with the increasing prevalence of Connected Automated Vehicles (CAVs) and vehicles outfitted with Advanced Driver Assistance Systems (ADAS). These advancements are pivotal in enhancing road safety, improving traffic efficiency, and reducing environmental impacts. Integral to this evolution is the deployment of various European Telecommunications Standards Institute (ETSI) services such as Cooperative Awareness Service (CAS), Collective Perception Service (CPS), Decentralized Environmental Notification Message (DENM), and Maneuver Coordination, which collectively contribute to a more interconnected and informed vehicular network.
The Collective Perception Service (CPS), in particular, stands out for its ability to mitigate blind spots by sharing locally sensed objects over IEEE 802.11p networks. This functionality significantly elevates environmental awareness, allowing vehicles to detect other vehicles, Vulnerable Road Users (VRUs), and other obstacles more effectively. However, the initial rollout of CPS is challenged by its low market penetration rate. Studies indicate that a minimum market penetration rate of 25\% is essential for CPS to achieve a nearly complete awareness ratio of surrounding VRUs, objects and vehicles in critical range, which is crucial for the proposed safety enhancement \cite{schiegg21}.
While DENM employs multi-hop message dissemination to extend its reach, such a mechanism is not currently utilized in CAS and CPS. This discrepancy is noteworthy, especially considering that multi-hop dissemination could be beneficial in scenarios of low CPS market penetration, thereby augmenting environmental awareness. Nevertheless, the implementation of multi-hop in CPS is constrained by the necessity for time-critical information dissemination and the maintenance of a low age of information to construct a real-time Local Environmental Model (LEM) for Intelligent Transport System (ITS) stations. Moreover, the potential for additional channel load poses another challenge to the default setting of multi-hop communication.
Given these considerations, our study focuses on exploring the impact of multi-hop communication within the CPS, particularly under conditions of low market penetration. We aim to determine whether multi-hop can effectively improve the total perception ratio in such scenarios. Furthermore, we investigate the potential advantages of multi-hop in CPS when there are underutilized channel capacities.
To this end, we propose an application layer-based approach for sharing perceived objects. An extension of the CPS is implemented, which decides whether an object received by other ITS-Ss will be forwarded. Through simulation of the CPS according to ETSI standards, this novel approach is compared against the baseline CPS performance and also juxtaposed with the Geographically-Scoped Broadcast (GBC) variant of ETSI GeoNetworking forwarding, which operates at the network layer. Through this comparative analysis, our study seeks to provide a comprehensive understanding of the effectiveness of multi-hop communication in CPS, especially in enhancing environmental awareness during the initial phases of CPS market penetration.

\section{Related Work}
\subsection{Multi-Hop in VANets}
In the realm of forwarding and multi-hop in Vehicular Ad-hoc Networks (VANets), recent research has made significant strides.
Forwarding V2X messages to improve awareness is not an entirely novel concept. In their research, Schmidt et al. \cite{5698243} introduce a message forwarding approach for beacons aimed at increasing vehicle awareness. Previous studies had primarily concentrated on expanding the awareness radius instead.
The study by \cite{pei2022} introduces optimized multi-hop broadcast schemes for emergency messages in vehicular networks, significantly enhancing the performance of Beacon and Emergency messages by reducing collision probabilities. This is achieved through the allocation of independent resource grants and the adjustment of forwarding nodes. Another work by \cite{sakai2023} demonstrates the feasibility of low-latency multi-hop communication using V2X Sidelink, with end-to-end latency maintained under 100 milliseconds even with three hops, a crucial factor for real-time vehicular applications. Further, the research presented in \cite{han2019} offers an adaptive scheme for network segmentation and channel allocation in large-scale V2X networks, effectively reducing multi-hop dissemination latency and improving network efficiency. Lastly, the study in \cite{vasa2022} explores the challenges in cooperative multi-hop platooning, particularly under varying propagation scenarios and system factors like inter-vehicle distance, highlighting the complexities in maintaining efficient V2X communication in dynamic environments. 

\subsection{Routing in Vehicular Communication}
\begin{figure}[t]
    \centering
    \vspace{0.25cm}
    \includegraphics[width=0.32\textwidth]{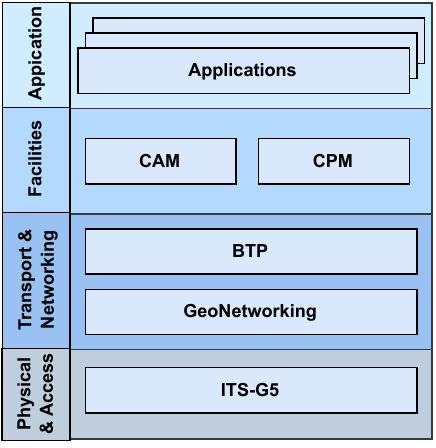}
    \caption{ETSI ITS-G5 Protocol Stack}
    \label{fig:its-g5l}
\end{figure}

Several techniques and services exist for enabling multi-hop communication in VANets to enhance the dissemination range of messages. Figure \ref{fig:its-g5l} shows the protocol stack proposed by ETSI for ITS-S. Implementing multi-hop at the network layer (Layer 3) of the OSI model is an obvious solution, however it comes with overheads such as latency, acknowledgements (ACKs), and processing delays that can reduce efficiency \cite{abbasi2019}. Efficient forwarding algorithms like the Intelligent Forwarding Protocol (IFP) aim to minimize these overheads by removing handshaking mechanisms and ACK dependencies. IFP achieves around 80\% lower end-to-end latency compared to traditional delay-based multi-hop protocols \cite{abbasi2019}.

While modifying network layer parameters can optimize multi-hop performance, it reduces compatibility with existing standards. The ETSI ITS protocol stack relies on GeoNetworking at Layer 3 for forwarding incoming messages between ITS stations (ITS-S) \cite{geo2017}. ITS services like Decentralized Environmental Notification Messages (DENM) implement multi-hop support at the application layer \cite{bellache2017_dcc}. Studies show that integrating a contention-based forwarding scheme for DENM with Decentralized Congestion Control (DCC) reduces channel congestion compared to non-contention based approaches \cite{bellache2017_contention}.

Since our study aims to maximize compatibility with existing ITS-Ss, no lower layer parameters are modified. However, it is valuable to survey prior multi-hop optimizations and explain how forwarding is handled in the ETSI ITS stack. IFP demonstrates the potential gains of optimizing handshakes, ACKs, and contention windows. Further research can investigate integrating such optimizations with standards-compliant designs above Layer 3.

\subsection{ETSI GeoNetworking}
GeoNetworking provides a set of broadcast services and forwarding algorithms to enable multi-hop distribution of packets in vehicular ad hoc networks. 
This section will motivate why GeoNetworking's versatile approach to ad hoc broadcasting and forwarding is well-suited as a platform for collective perception and cooperative vehicular applications.

The ETSI GeoNetworking protocol \cite{geo2017} specifies several broadcast services for multi-hop distribution of packets in an ad hoc network. Single Hop Broadcast (SHB) provides simple direct neighbor delivery without forwarding. Topologically-Scoped Broadcast (TSB) forwards packets up to a configured maximum hop count. It can use greedy forwarding, where each node selects its neighbor closest to the destination as the next forwarder. TSB can also use contention-based forwarding, where nodes broadcast the packet and receiving nodes set a timer to rebroadcast that is inversely proportional to their distance from the previous sender. Hearing a duplicate cancels the timer.

GeoNetworking also defines Geographically-Scoped Broadcast (GBC) to constrain distribution to a geographic area. Within the target area, GBC can flood packets via simple rebroadcasting. It can also use the contention-based forwarding algorithm described above. An advanced forwarding method combines greedy selection of a next hop neighbor with enhancements to contention-based forwarding such as sectorized rebroadcasting areas and controlled retransmissions for efficiency and reliability. Outside the target area, GBC uses line forwarding with greedy or contention-based forwarding to route packets towards the area boundary. The related Geographically-Scoped Anycast service delivers packets to a single node within a geographic area.

Greedy forwarding provides optimal forwarding selection but lacks redundancy. Contention-based methods enable reliability through duplicate transmissions and timer coordination between nodes, at the cost of duplicates. Advanced forwarding tune parameters such as rebroadcast sectors and transmission limits to improve performance while retaining useful contention-based behavior.

\subsection{Collective Perception}
In the field of Cooperative Awareness and Collective Perception, approaches exist to forward messages via Road Side Units (RSUs). Researchers found that the relay and aggregation of CAMs in a RSU help to reduce the Time To Arrival (TTA) of vehicles approaching the intersection, thus increasing the traffic flow\cite{Garlichs20}.
A Collective Perception study focusing on VRU awareness found that the deployment of a central RSU at intersections leads to increased VRU perception, while maintaining a moderate channel load. The CPS implemented in the RSU decided which objects are forwarded \cite{vehits23}.  
However, there is a lack of studies focusing on decentralized forwarding techniques, involving all ITS-Ss of a VANet, with emphasis on low market penetration at initial CPS rollout. 
\begin{figure}[t]
    \centering
    \includegraphics[width=0.49\textwidth]{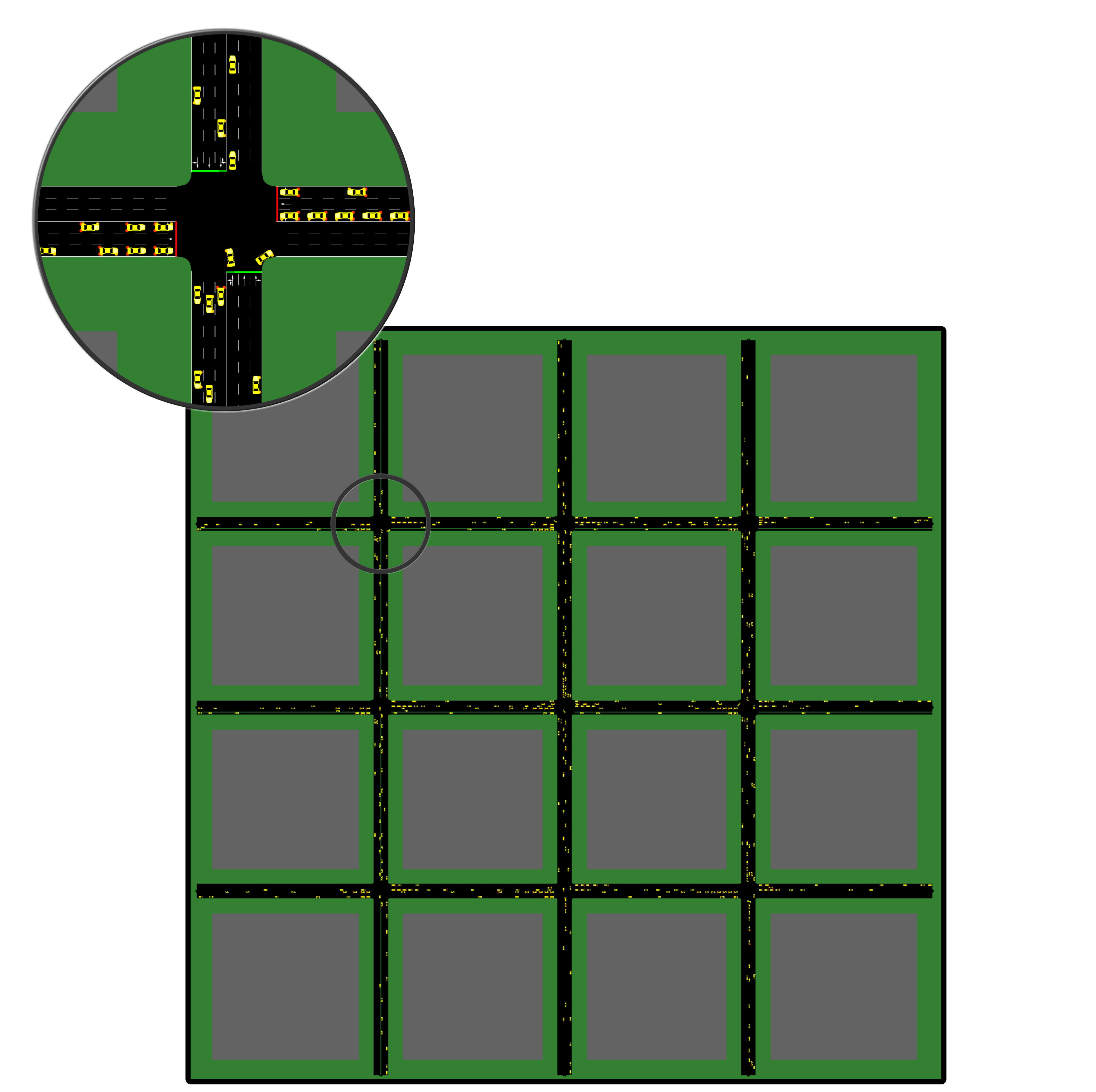}
    \caption{Simulated SUMO scenario}
    \label{fig:sumo}
\end{figure}
\section{Contribution}
As stated in the previous section, several research exist which study multi-hop forwarding and its application, i.e. extending the range of the DEN service \cite{bellache2017_dcc}.
However, we identify a lack of studies in the realm of Collective Perception, in which multi-hop message forwarding could be beneficial to extend the environmental awareness of ITS-Ss. At first glance, it may be obvious that the implementation of multi-hop would reach more recipients, and therefore improve the environmental perception.
However, forwarding all the messages in an uncontrolled fashion may quickly saturate the communication channel, leading to an overall deterioration of the service.

In this work, we aim to address several open research questions , which are noteworthy. Firstly, we implement an application layer based forwarding algorithm and evaluate its performance in comparison to a routing-based forwarding algorithm. Secondly, we examine the magnitude of additional channel load resulting from multi-hop messages and whether the consequent channel congestion is a reasonable trade-off. Thirdly, we investigate the benefit in terms of safety metrics, particularly focusing on the Environmental Awareness Ratio (EAR). Lastly, we assess to what extent the Age of Information (AOI) is increased by implementing multi-hop algorithms and determine if the AOI is sufficient for real-time applications.

To achieve comparability of results to other research, we stick to ETSI standards, implementing the protocol stack (shown in Figure \ref{fig:its-g5l}). State of the art V2X and mobility frameworks are used to simulate realistic vehicular dynamics and communication, as stated in section \ref{sec:sim}.
\begin{figure}[t]
\vspace{-0.5em}
\begin{algorithm}[H]
\label{algorithm1}
\caption{Collective Perception Message Forwarding}
\begin{algorithmic}[1]
\State Initialize Local Environment Model (LEM): $LEM \gets \emptyset$
\State Initialize CPM Buffer: $CPM\_Buffer \gets \emptyset$
\State Define maximum hop count: $MAX\_HOP\_COUNT$

\While{true}
    \State PERIODIC\_EXECUTION\_WAIT()
    \While{not $CPM\_Buffer$.isEmpty()}
        \State $CPM \gets CPM\_Buffer$.dequeue()
        \For{each object in $CPM.PerceivedObjectContainer$}
            \State $objID \gets \text{object.id}$
            \If{not $LEM[objID]$ OR ($LEM[objID].\text{timestamp} < \text{object.timestamp}$ AND $\text{object.hopCount} < MAX\_HOP\_COUNT$)}
                \State $LEM[objID] \gets \text{object}$
            \EndIf
        \EndFor
    \EndWhile

    \State $newCPM \gets$ CREATE\_NEW\_CPM()
    \For{each object in $LEM$}
        \If{($\text{object.kinematicChangeTrigger()}$) AND 
            $\text{object.hopCount} < MAX\_HOP\_COUNT$}
            \State $newCPM.\text{add}(\text{object})$
        \EndIf
    \EndFor

    \If{DCC\_SEND\_CONDITION()}
        \State SEND\_CPM($newCPM$)
    \EndIf
\EndWhile
\end{algorithmic}
\end{algorithm}
\vspace{-2em}
\end{figure}

\section{Application Layer Forwarding}
\label{sec:application_layer_fowarding}
Algorithm 1 shows the principle for the application layer based forwarding, which is integrated into the CPS on all ITS-S and later compared to the baseline CPS approach, defined by ETSI. The algorithm operates by maintaining a Local Environment Model (LEM), which is periodically updated with data from received Collective Perception Messages (CPMs) buffered during inter-execution intervals, expressed by $PERIODIC\_EXECUTION\_WAIT()$. The waiting period equals the lowest interval between the generation of two CPMs, which is set to 100 ms. The core of the algorithm lies in its ability to discern and integrate only the most recent and relevant object data into the LEM. This is achieved by comparing the timestamps of incoming object data against those already stored in the LEM, ensuring that only newer information is retained. Additionally, the algorithm incorporates a hop count-based filtering mechanism to ensure that the received objects in CPMs are not excessively forwarded. In our simulation, $MAX\_HOP\_COUNT$ is set to 2. It is important to consider that the LEM does also contain local sensed objects. Therefore the resulting CPM consists of objects that are being forwarded and objects that are detected by the ego vehicle. Objects are included in the newly generated CPMs if they meet specific criteria, according to the ETSI generation rules defined in \cite{CPS2023} ($object.kinematicChangeTrigger()$): a change in position exceeding a distance of 4 meters, a speed change surpassing 4 meters per second, a heading change over 4 degrees, or a lapse of more than 1 second since their last inclusion in a CPM, provided their hop count is below the predefined maximum threshold. This ensures that the generated CPMs are not only up-to-date but also relevant and concise. The decision to transmit the CPM is contingent upon the approval of the Decentralized Congestion Control (DCC) mechanism, aligning the frequency of CPM dissemination with channel load requirements.
\begin{table}[t]
\vspace{0.5em}
    \centering
        \normalsize
    \caption{Simulation parameters}
    \footnotesize
    \begin{tabularx}{0.45\textwidth}{l|l}
         \textbf{Parameter} & \textbf{Value}  \\
         \hline
         Enabled services & ETSI CPS\\
         \hline
         Physical layer & ITS-G5\\
         \hline
         Bit rate & 6Mbit/s \\
         \hline 
         Carrier frequency & 5.9GHz \\
         \hline
         Bandwidth & 10MHz \\
         \hline
         Channel & G5-CCH \\
         \hline
         Transmission power & 200 mW \\
         \hline
         Signal threshold & -85 dBm \\
         \hline
         Noise threshold & -65 dBm \\
         \hline
         Signal propagation model & GEMV2
         \\
         \hline
         Vehicle sensor set & 360\textdegree camera, 85m range\\
         \hline
         Vehicle densities & 30 Veh/km, 60Veh/km\\
         \hline
         Penetration rates & 05\%, 10\%, 25\%, 50\% \\
         \hline
         Simulation time & 10 runs, 15 s per run\\
         \hline
         Decentralized Congestion Control & Enabled, reactive approach\\
         \hline

    \end{tabularx}
    \label{tab:sim}
\end{table}

\section{Simulation Setup}
\label{sec:sim}
In our study, we utilize the Artery framework to simulate Vehicle-to-Vehicle (V2V) communication \cite{artery2015}, adhering to the ETSI ITS-G5 standards \cite{etsi_itsg5}. This simulation is integrated with the Simulation of Urban MObility (SUMO) traffic simulator \cite{dlr127994} to create an urban traffic scenario. A screenshot is provided in figure \ref{fig:sumo}. The scenario is generated based on a Manhattan grid layout, featuring roads with two lanes in each direction, to replicate typical urban traffic conditions. The grid size is 3x3, with 9 intersections in total. The total grid size is 1km x 1km. Two distinct vehicle densities are modeled: a low density of 30 vehicles per kilometer and a high density of 60 vehicles per kilometer. To accurately represent the challenges in V2V communication, obstacles are strategically placed at each intersection, serving to simulate the shadowing effect commonly encountered in dense urban environments. Table \ref{tab:sim} lists the simulation parameters. The ETSI ITS-G5 standard is implemented on lower layer. For a more realistic portrayal of shadowing in our model, the Geometry-based Efficient propagation Model for V2V communication (GemV2) is enabled. Our simulation strictly conforms to ITS-G5 and ETSI standards, ensuring its applicability and relevance to real-world scenarios. Additionally, Decentralized Congestion Control (DCC) mechanisms are activated to control the frequency of CPM dissemination. The reactive DCC is applied, according to \cite{etsi2018}. Logging is enabled for all vehicles located within the central area of 900 meters by 900 meters. The vehicles are equipped with a 360 degree camera sensor, with a radius of 85 m, derived from the model described in \cite{9922147}.
For the evaluation, three different CPS modes are tested and analyzed: As a baseline comparison, the ETSI CPS without forwarding is applied. The second mode is the baseline CPS with enabled Geographically-Scoped Broadcast (GBC) on network layer. To avoid extensive network overload, we set the maximum lifetime of messages to 1 s and the maximum radius of 200 m, in which messages will be forwarded, based on the source ITS-S of the message. A range of 200 m ensures that the entire approaching intersection arm is covered for each vehicle. Analogous to the application layer algorithm, the GBC maximum hop limit is also set to 2. The third tested mode is the application layer based algorithm described in \ref{sec:application_layer_fowarding}.

\section{Results}
\begin{figure}[t]
  \centering
  \begin{subfigure}[b]{0.48\textwidth}
    \includegraphics[width=\textwidth]{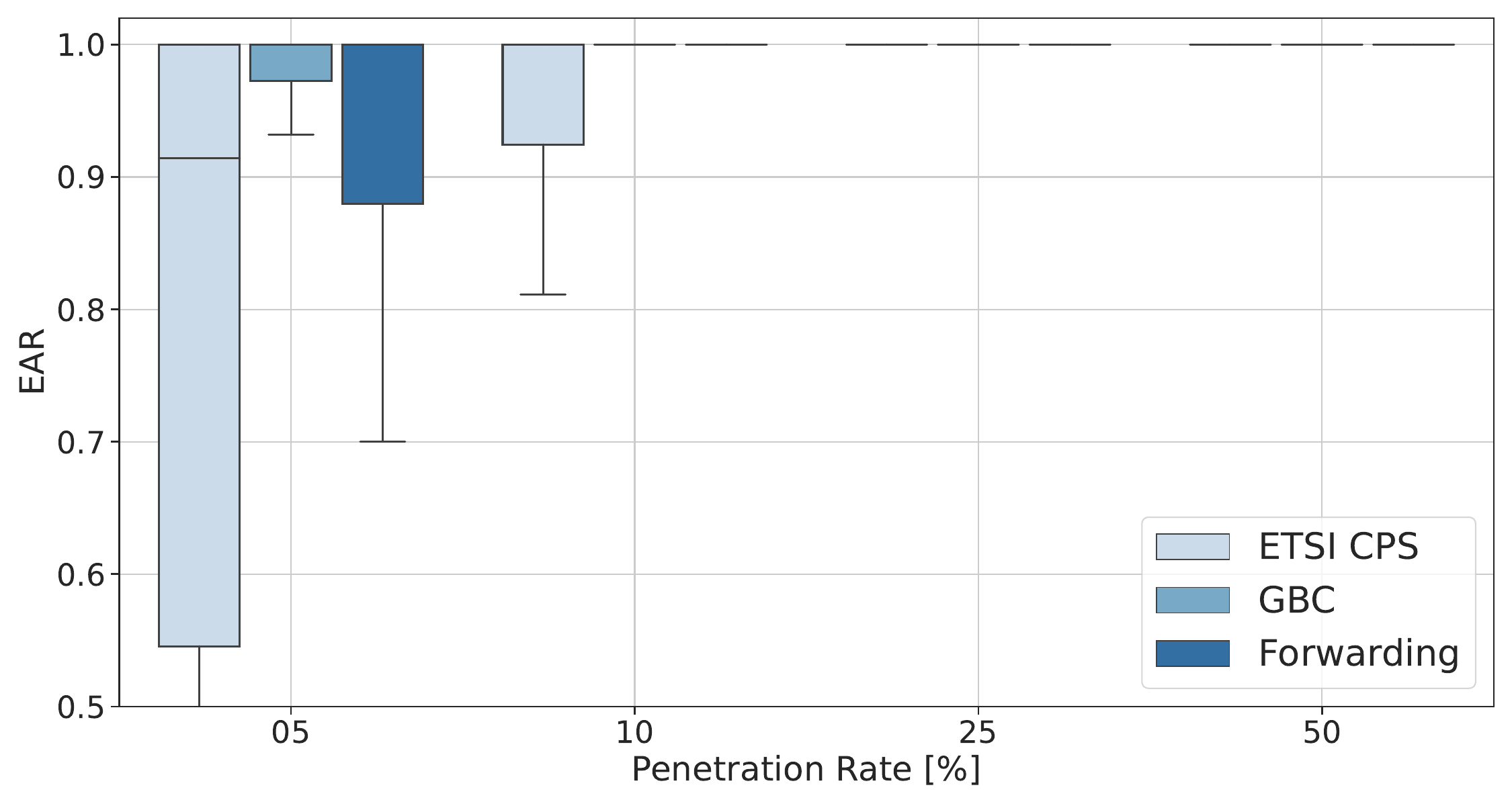}
    \caption{Low Traffic Density}
    \label{fig:ear_low}
  \end{subfigure}
  \begin{subfigure}[b]{0.48\textwidth}
    \includegraphics[width=\textwidth]{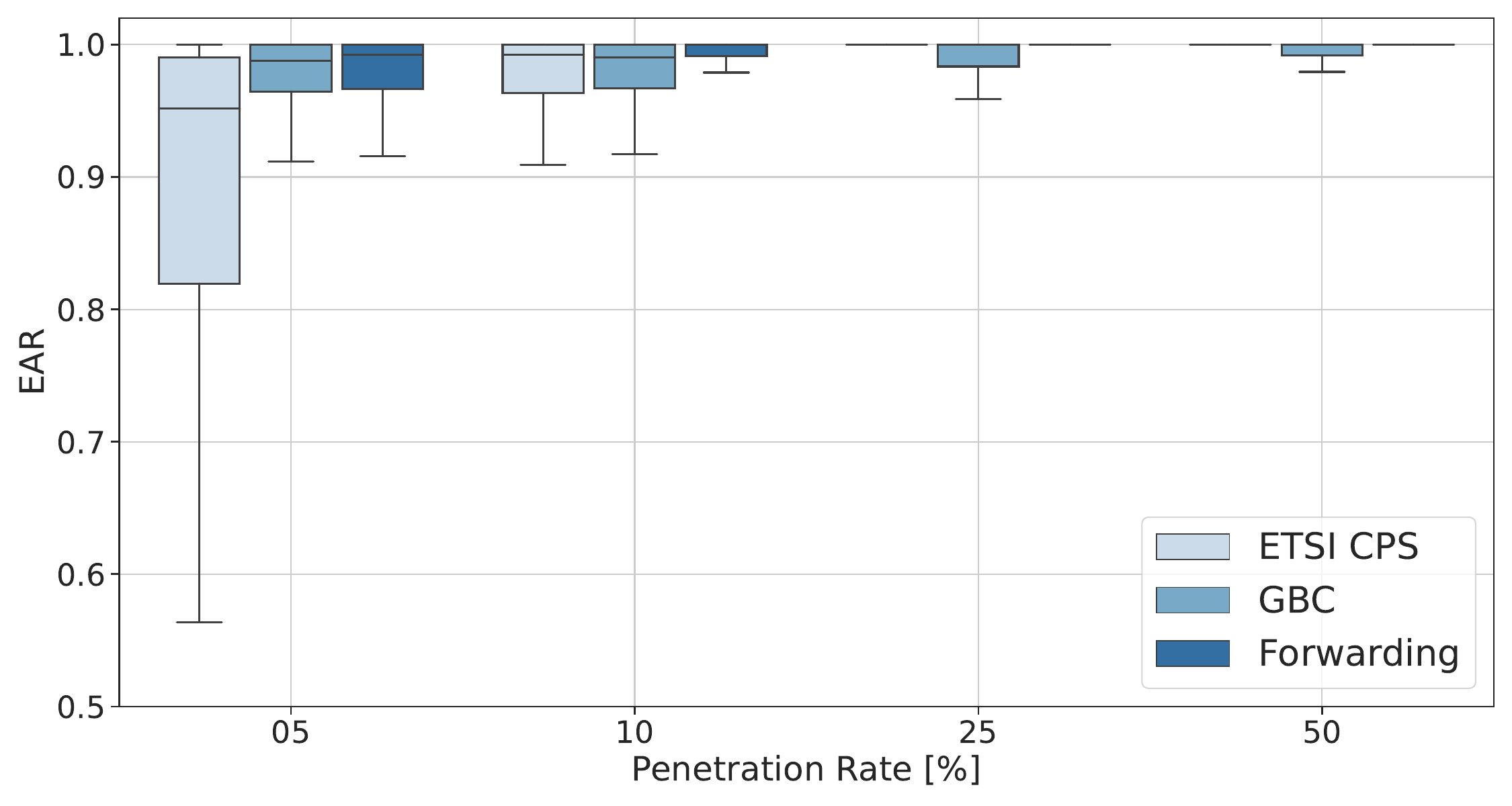}
    \caption{High Traffic Density}
    \label{fig:ear_high}
  \end{subfigure}
  \caption{Environmental Awareness Ratio for different modes, low and high traffic density scenario}
  \label{fig:ear}
\end{figure}
\begin{figure}[t]
    \centering
    \includegraphics[width=0.48\textwidth]{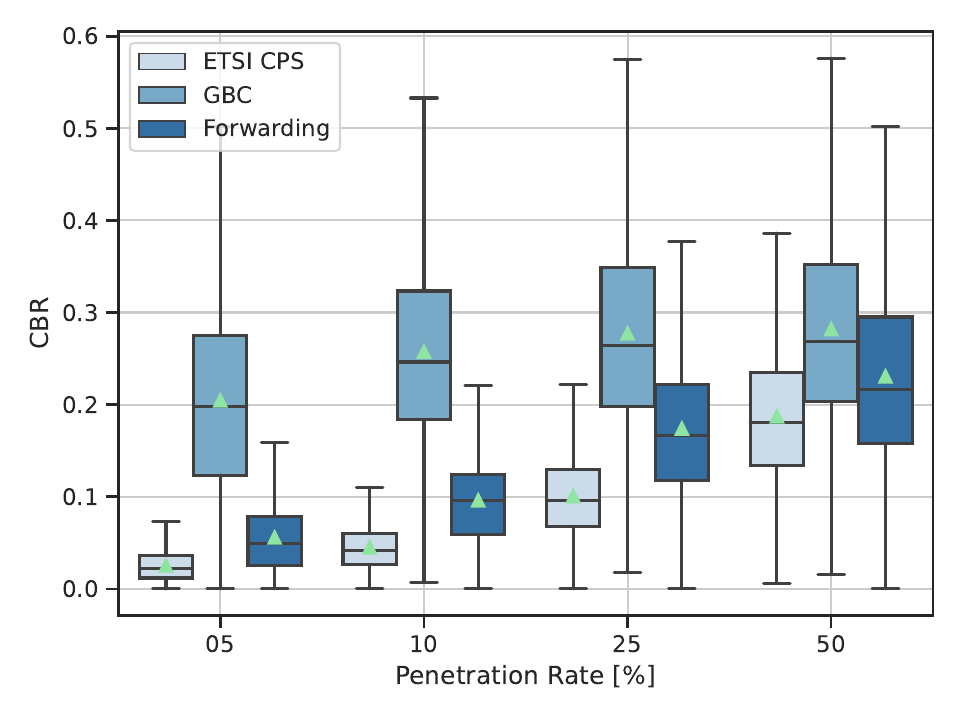}
    \caption{CBR for high traffic density scenario}
    \label{fig:cbr}
\end{figure}

Metrics on both the network and application layers were examined to assess the impact of our forwarding algorithm on the performance of Collective Perception. We compared three distinct Collective Perception Service (CPS) modes: ETSI CPM, no forwarding, application layer forwarding, and GBC forwarding. To evaluate the impact on ITS-Ss' environmental awareness, we utilized the Environmental Awareness Ratio (EAR) \cite{schiegg21}, depicted in Figure \ref{fig:ear}. It computes the ratio of perceived objects to the sum of objects in the complete range of interest. Perceived objects are defined as all objects received by V2X communication or sensed by local sensors. The radius of the range of interest is set to 200 m, and the maximum Age Of Information (AOI) to 1 s. This metric was analyzed for both low and high traffic densities. An expected trend was observed: the EAR increased with the penetration rate for both vehicle densities across all CPS dissemination modes. Notably, significant variations were seen among the three modes at fixed penetration rates. In low density scenarios, the median EAR for ETSI CPS mode was 0.92, while the medians for GBC and forwarding reached 1 at a mere 5\% penetration rate. The lower quartiles revealed even starker differences: the EAR for ETSI CPS was 0.56, improving to 0.98 (GBC) and 0.89 (forwarding). At a 10\% penetration rate, both GBC and forwarding achieved 100\% EAR, whereas ETSI CPS remained at 0.92. At higher penetration rates, the performance of all three modes was comparable. In high-density case, similar shifts in EAR were observed; however, a 100\% EAR was not attained for all three modes. With increasing penetration rates, GBC mode fell behind, failing to achieve a 100\% EAR in the lower quartile and whisker.

For low vehicular density, we observed a slight outperformance of GBC over forwarding, which was not replicated in high vehicle density scenarios. In these, our application layer-based forwarding algorithm led the performance. Further analysis on the impact of these modes on network load was conducted using the Channel Busy Ratio (CBR), shown in Figure \ref{fig:cbr} for high traffic density, chosen specifically to illustrate the upper bound of CBR under such conditions. The CBR, dependent on \( T_{\text{ON}} \) and \( T_{\text{OFF}} \), is the proportion of time a communication channel is actively used for transmission (\( T_{\text{ON}} \)) compared to the total time, which includes both transmission (\( T_{\text{ON}} \)) and idle periods (\( T_{\text{OFF}} \)).
GBC mode consistently exhibited the highest CBR across all penetration rates. The disparity between the modes was pronounced at low penetration rates, but less so with increasing rates. At 25\% and 50\% penetration, channel saturation was evident in GBC mode, attributed to the Dynamic Channel Access (DCC) mechanism's sending frequency rate throttling, ensuring moderate channel load. The forwarding mode achieved significantly lower CBR at all penetration rates compared to GBC. However, the baseline ETSI CPS had the lowest CBR, as expected due to the absence of forwarding.
\begin{figure}[t]
    \centering
    \includegraphics[width=0.48\textwidth]{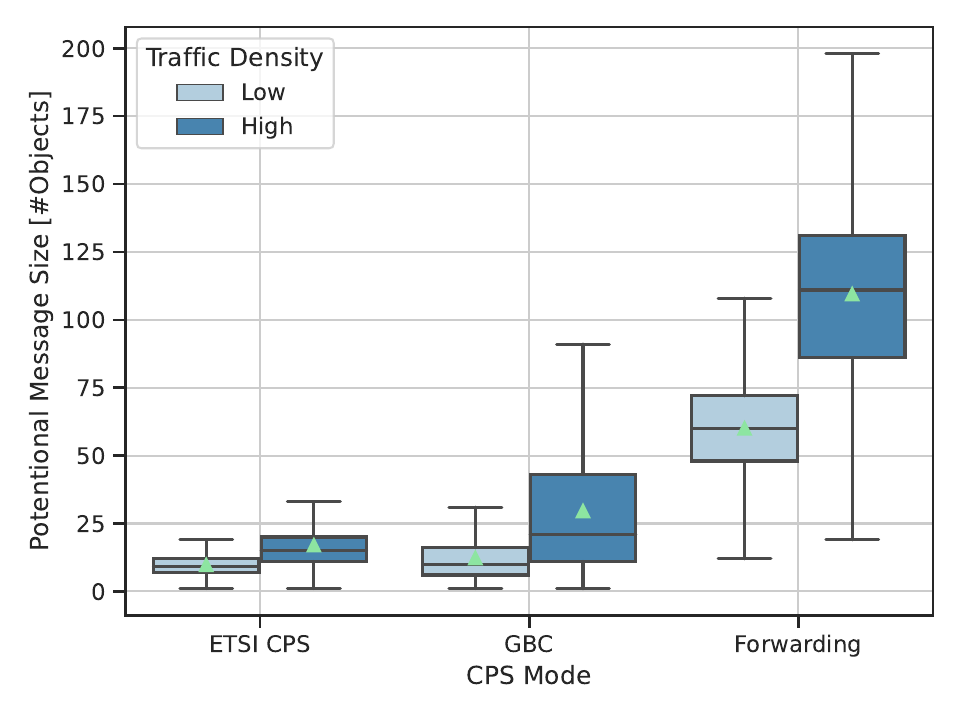}
    \caption{Potential number of objects included in a CPM, 25\% penetration rate}
    \label{fig:cpm_size}
    \vspace{-1.4em}
\end{figure}
To point out the differences between modes in greater detail, we examined the potential number of objects in the PerceivedObjectContainer of the CPM, illustrated in Figure \ref{fig:cpm_size}. This metric reflects the number of objects that would be sent in a single CPM, barring constraints imposed by CPS standards (limiting to 128 objects per message) and fragmentation. It shows the number of objects, the specific algorithm would choose to include in one CPM. For the analysis, we set the penetration rate to 25\% in order to show the differences for the tested modes for one typical scenario. The application layer-based forwarding exhibited the largest container size in both high and low traffic density scenarios, resulting in reduced overhead per CPM due to an increasing payload-to-packet header ratio. Interestingly, in GBC mode, which adheres to the same rules as ETSI CPS, the message size increased. This can be correlated with the CBR, as shown in Figure \ref{fig:cbr}, where channel saturation begins at a 25\% penetration rate, leading to message throttling due to DCC. Consequently, more objects are sent per CPM as the dissemination frequency decreases, fulfilling the kinematic object inclusion criteria set by the ETSI standard.

\begin{figure}
    \centering
    \includegraphics[width=0.48\textwidth]{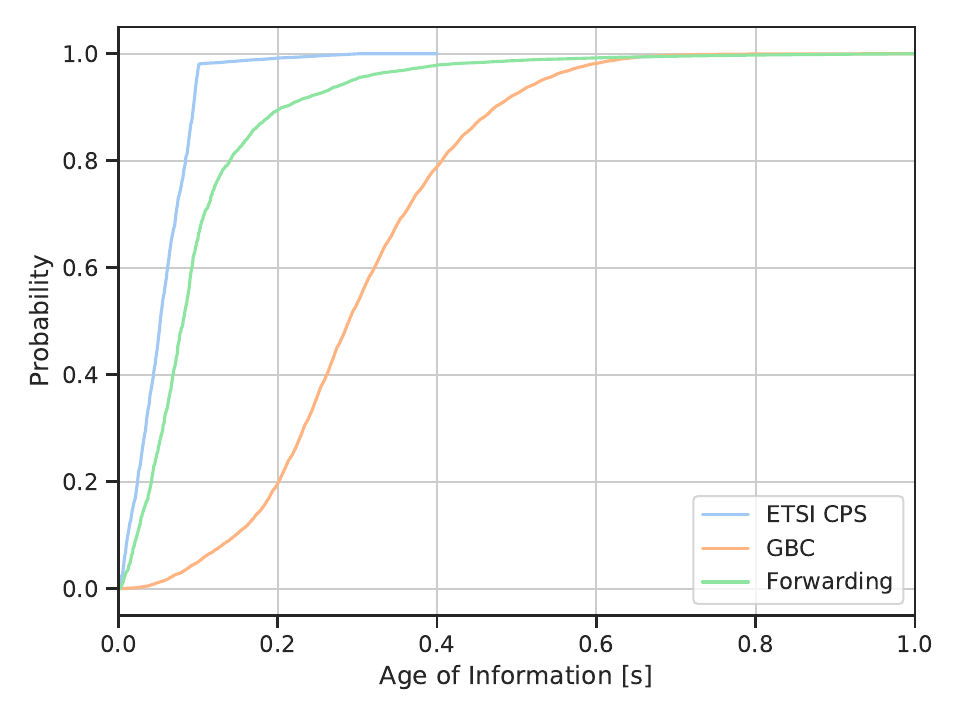}
    \caption{Age of information}
    \label{fig:aoi}
\end{figure}

Figure \ref{fig:aoi} illustrates the Age of Information (AOI) of the received objects as a Cumulative Distribution Function (CDF). For this evaluation, the penetration rate was set to 25\% and results for both vehicle densities were aggregated. The AOI, defined as the time difference between when the sending vehicle conducts its sensor measurement and the time of reception by another ITS-S, revealed expected trends. No forwarding resulted in the lowest AOI, with a 99th percentile of less than 180 ms. Our forwarding algorithm delivered 85\% of information within 200 ms, significantly outperforming the GBC mode, where less than 20\% of objects were under 200 ms old. The median AOI for GBC was approximately 285 ms, compared to 55 ms for ETSI CPS and 80 ms for the application layer-based forwarding mode.


\section{Conclusion}
We show a proposal for an application layer based implementation for an application of multi-hop Collective Perception. In our simulation according to ETSI standards, the implementation outperforms ETSI CPM in low market penetration rate scenarios, providing a high perception ratio while maintaining moderate channel load. We further show that our implementation outperforms of geographically-scoped forwarding (GBC) implementation, by providing a similar environmental awareness ratio with significant less channel congestion. The Age of Information is increased compared to the baseline ETSI CPS, however the median AOI remains below 80 ms, making real-time perception applications such as the CPS possible. In future research, the parameters of the algorithm can further be optimized, depending on certain application requirements, as the trade-off between channel congestion and perception ratio. In the context of an urban scenario test, it would be a logical and valuable extension to incorporate VRUs to demonstrate the forwarding performance in relation to VRU perception.

\section*{Acknowledgment} This publication is funded by the Lower Saxony Ministry of Science and Culture under grant number ZN3493 within the Lower Saxony “Vorab“ of the Volkswagen Foundation and supported by the Center for Digital Innovations (ZDIN).

\bibliographystyle{ieeetr}
\bibliography{bibliography}

\end{document}